\documentclass[12pt,letterpaper]{article}
\usepackage{ifthen}

\newcommand{\tv}[1]{\ensuremath{\mathbf{#1}}}

   \newcommand{\dimenv}[2]{\ifthenelse{#2=1}
         {\ensuremath{[\mathrm{#1}]}}
         {\ensuremath{[\mathrm{#1}]^{#2}}}}

   \newcommand{\unitenv}[2]{\ifthenelse{#2=1}
         {\ensuremath{\mathrm{#1}}}
         {\ensuremath{\mathrm{#1}^{#2}}}}

   \newcommand{\abs}[1]{\left|{#1}\right|}

   \newcommand{\deriv}[3]{
	\ifthenelse{\equal{#3}{1}}
		{\ensuremath{\frac{\mathrm{d}{#1}}{\mathrm{d}{#2}}}}
		{\ensuremath{\frac{\mathrm{d}^{#3}{#1}}{\mathrm{d}{#2}^{#3}}}}
	}

\newlength{\sectionskip}
\renewcommand\section[1]{\vspace{\sectionskip}\pagebreak[2]\par%
  \refstepcounter{section}\noindent\textbf{\thesection.\ {#1}}%
  \nopagebreak\par%
  \nopagebreak\vspace{0.5\sectionskip}\nopagebreak}

\newcounter{problem}
\newcommand{\problemname}{Problem}
\newcommand{\problem}[1]{\refstepcounter{problem}%
  \noindent\textbf{{\problemname~\theproblem}:\ }\textsl{{#1}}%
  \par\vspace{0.5\sectionskip}}

\newcommand{\titleandetc}[2]{\par\noindent\textbf{#1}\footnote{#2}}
\newcommand{\authorandetc}[3]{\vspace{\sectionskip}\par\noindent{#1}%
  \par\noindent{#2}\par\noindent{#3}}
\renewenvironment{abstract}{\vspace{\sectionskip}\small\par\noindent%
  \textbf{Abstract:} }{\vspace{\sectionskip}\par}

\begin{document}

\titleandetc{Real-world ballistics: A dropped bucket}%
  {Copyright 2007 David~W.~Hogg (david.hogg@nyu.edu).  You may copy
  and distribute this document provided that you make no changes to it
  whatsoever.  It is a pleasure to thank Sanjoy Mahajan and Greg
  McDonald for helpful comments on various drafts.}%
\authorandetc{David W. Hogg}%
             {Department of Physics, New York University}%
             {2007 September 1}

\begin{abstract}
I discuss an apparently simple ballistics problem: the time it takes
an object to fall a small vertical distance near the surface of the
Earth.  It turns out to be not so simple; I spend a great deal of time
on the quantitative assessment of the assumptions involved, especially
with regards to the influence of the air.  The point is \emph{not} to
solve the problem; indeed I don't even end up solving the problem
exactly.  I introduce dimensional analysis to perform all of the
calculations approximately.  The principal theme of the lecture is
that \emph{real} physics can be very different from ``textbook''
physics, since in the real world you aren't ever told what equations
are appropriate, or why.
\end{abstract}

I was walking to work one morning and above me on the third story of a
scaffolding there was a man washing windows.  As I was walking under
him, he accidentally knocked his bucket and the bucket started to fall
from the scaffolding towards me\footnote{Story may be apocryphal.}.
How much time did I have to jump out of the way?

\section{Real physics, not textbook physics}

I expect that almost anyone reading this has had some physics
instruction of some kind in secondary school or from equivalent books
or classes.  I want to emphasize that the ``physics'' one learns in
these contexts might not be all that useful or relevant to
understanding the \emph{real} physics we are going to discuss here and
in the lectures that follow.  This is true even for---maybe
\emph{especially} for---students especially \emph{good} at that kind
of ``textbook'' physics.  Real physics is very different from textbook
physics.

Doing real physics involves intuition, approximation, and quantitative
reasoning to understand what \emph{matters} in a physical situation.
We use formal techniques (such as calculus and proofs) only once we
have understood the fundamental properties of the physics problem at
hand.  There are \emph{scalings} we can find, which tell us how the
problem solution depends, functionally, on each input parameter.  We
can analyze \emph{limiting behavior} of the system when it is
simplified in various ways, many of which are physically illuminating.
We can determine, quantitatively, the \emph{dominant processes,} the
physical effects or forces that make the biggest contributions to the
system's behavior\footnote{If we say ``dominant'' we usually mean
not just the biggest, but so much bigger that we can ignore the
others.}.  We are rarely concerned with the particulars of an exact
solution to a well-defined mathematically solveable problem, although
sometimes one will emerge beautifully after our---much more
important---rough work on the problems.

In contrast, in textbook physics, a lot of what you do involves
equations: how to use and derive them.  The equations are related to
physics, of course, but the focus is on the equations themselves.  You
probably remember many of them: ``$v_0\,t+(1/2)\,a\,t^2$,''
``$(1/2)\,m\,v^2$,'' ``$F=m\,a$.''  You also solve many problems in a
textbook class, but those problems are tuned carefully to make use of
the very equations you had been given.  They are not physics
questions, really, they are cleverly disguised mathematical ``word
problems.''  Textbook physics classes teach techniques for solving
those problems---problems well-matched to the store of equations.  In
these lectures we not going to concentrate on those textbook skills,
though we \emph{will} use some of them.

Don't get me wrong: We \emph{are} going to solve problems.  We are
going to use equations and some beautiful mathematics.  But we are not
going to make our \emph{goal} the mathematical solutions of those
particular problems.  The problems we do---and we will do several per
lecture---will be done with the over-riding goal of making ourselves
better physicists; better at \emph{real} physics.

Let us return to the problem: I am under the window-washer's platform
and the bucket has begun to fall.  What should I do?  There is a clear
textbook ``strategy'' for the solution of this problem: Find the
relevant equation (from the book), plug in numbers, and get an answer.
There are two problems with this strategy: \textsl{(1)}~In the real
world, no-one tells you what equations to use, and
\textsl{(2)}~scaffolding and buckets aren't labeled with their heights
and masses and contents.  In other words, the problem is ill-posed.  I
haven't told you what to assume, physically, and I haven't told you
what numbers to plug in.

In the real world, \emph{all questions of importance are ill-posed.}
An ill-posed problem is ``explain the motions of the planets.'' A
well-posed problem is ``compute the relative semi-major axes of Mars
and the Earth, using the tabulated sidereal periods of the planets and
Kepler's laws.''  An ill-posed problem is ``can we generate our
electricity with wind power?''  A well-posed problem is ``what is the
torque on a propeller with blades of a certain length, width, pitch
angle, and cross-sectional shape, given a steady wind of a certain
speed?''  Well-posed questions can only be asked once significant
progress has been made on the ill-posed questions.\footnote{A similar
point is well made in Mazur,~E., ``The problem with problems,''
\textit{Optics \& Photonics News,} \textbf{6} 59--60 (June 1996).}

This bucket problem is ill-posed, but is it \emph{important?}  It has
the great virtue that---despite being ill-posed---it has a well-posed
form and an exact solution under certain conditions (to be explored
below), so we can use it to test our methods.  It has this exact
solution, and it shares with important ill-posed problems the
requirement that we need to figure out the relevant physics on our
own; this makes it a good problem with which to start.  In
general---in these lectures and in the world---we have to figure out
what matters; we aren't going to be told.  If we are lucky, the
dominant physical effect will be something we \emph{can} represent,
approximately, with an equation.  If we can, we will derive that
equation.  Then we are going to use things we know about the world
(scaffolding, buildings, buckets, and the Earth) to figure out what
numerical quantities to insert into the variables of the equation we
derived.

I am spending a lot of time on this polemic, but for a good reason:
Most of what you have learned about physics up to this point is
\emph{highly misleading}.  I have just mentioned that ``if we are
lucky'' we will be able to represent the dominant physical effect with
an equation.  \emph{Most} questions you can ask about the physical
world do \emph{not} have an answer of this kind; most things you
observe have \emph{no good} description in terms of an equation you
can write down.  For example, if a window-washer drops not a bucket
but the front page of a broadsheet newspaper, there is no equation we
can write down that even gives a good approximation of the
\emph{probability} that the paper hits me (depending, as it does, on
the distribution of eddies and breezes in the air between us), let
alone how long takes.  If the window-washer drops the bucket not from
the third story but from the tenth, there are approximate equations we
can use, but if we want an answer accurate at the few-percent level,
we require a computer.

Most textbook presentations of physics you have seen previously have
``led you down the garden path'' in which have been planted simple,
solveable problems, described by simple, solveable equations.  In the
universe of all physics problems, the problems in this garden comprise
the exception, not the rule.  Worse, some presentations of physics
have \emph{inappropriately} applied simple, solveable equations to
situations where they are physically inapplicable, even approximately.
What we are going to do differently here is that we are going to spend
our time not solving or \emph{using} the equations, but we are going
to spend our time \emph{discovering} and \emph{interrogating} them.
To use an equation is one thing, to show that it is the appropriate
equation is quite another.

\section{The ingredients of a solution}

A real solution to a physics problem involves interrogating your
physical intuition to make some kind of guess or prediction, figuring
out all of the possibly relevant physical effects, figuring out which
are most important, making the necessary approximations to get an
equation or other mathematical representation, doing the math,
comparing your answer to your intuitive guess, and then finally
checking quantitatively your answer and your approximations using
non-trivial diagnostics.  The making of approximations is what turns
the ill-posed problem into a well-posed problem.  Exact formulae are
only used at the point that there is a well-posed problem, and they
comprise only the \emph{tiniest part} of this kind of complete
solution---a \emph{real} solution---of a physics problem.

So once again: The bucket is starting to fall.  Imagine it.  How long
do you think it will take to fall?  Will it happen in the blink of an
eye?  Will I watch it fall, and have time to think about it, perhaps
time to jump out of the way?  Or will it take many seconds, such that
I could have a substantial conversation with a friend as it falls?  In
general, when a physics problem describes a situation that you have
actually \emph{seen,} it is a good idea to use that experience.  So
take a moment to visualize this, and predict an answer, just from your
memory of similar events.

An important theme of these lectures is that fundamental
prin\-ci\-ples---fundamental properties of the Universe---appear in
all sorts of situations around you.

What can matter, physically, in the problem?  One quantity that
clearly matters is the vertical \emph{distance} $h$ that the bucket is
going to fall.  I think we would all agree that the larger the
distance, the longer the fall, or that time increases with distance.
What is the height of a third-story window?  Well, stories of
buildings are about $3\,\unitenv{m}1$, so the third story has a height
$h\approx 10\,\unitenv{m}1$.  You might imagine that it matters
whether the bucket falls from the floor level or window level; in
principle that matters, but at the level of precision of this
calculation it won't.  Don't take it from me, however: We will check
this at the end.  This is another theme that will recur in this
lectures and the lectures to follow: There is no need to determine any
numbers to better precision than the calculation supports.  With a
question as ill-posed as this one, we certainly don't need to know the
height $h$ with great accuracy.

What else can matter, physically, in the problem?  Clearly the size,
mass, contents, and composition of the bucket all matter.

Or do they?  A little knowledge can be a very dangerous thing.  From a
textbook-physics setting, you may recall the observation,
attributed\footnote{I say ``attributed'' here because I am not a
  historian; I don't want to assert anything historical.  However, from
  here on, I \emph{will} attribute this to Galileo.} to Galileo, that
all things fall at the same rate.  That is, the size, mass, contents,
and composition of the bucket \emph{can't matter}.

Or can they?  Galileo's observation is true only in a very certain
\emph{regime}, or set of physical conditions.  It is true in the
regime in which \emph{the air does not matter.}  Here are two examples
of situations in which the air \emph{does} matter: In one, the window
washer drops not a bucket but a page from a broadsheet newspaper.
That sheet will flip and flop and fold and drift down, because its
interaction with the air is just as important as the gravitational
force.  In another, the bucket is not a tin bucket filled with water,
but a rubber balloon filled with helium gas.  In this case, when the
window washer ``drops'' (releases?)\ it, the bucket falls not
downwards, but \emph{upwards}; for a helium balloon, the buoyant force
exceeds the direct gravitational force.

How can we tell whether we are in the air-doesn't-matter regime, and
how does that depend on the mass, size, contents, and composition of
the bucket?  Those who cannot answer that question, quantitatively,
have \emph{never really understood} Galileo's famous observation.  Use
of Galileo's result without the ability to quantitatively test its
applicability constitutes only textbook physics.  The purpose of these
lectures will be to develop the real, \emph{physical} understanding.

Our problem is ill-posed, so we \emph{don't know} the size, mass,
contents, or composition of the bucket for sure.  But, in fact, we all
\emph{do} have very good ideas of these, from our own experiences of
buckets.  A typical bucket holds about a cubic foot (which, since a
foot is about $1/3\,\unitenv{m}1$, is about $0.03\,\unitenv{m}3$), and if it is
filled with water, its mass is dominated by the water.  Water has a
density of $1000\,\unitenv{kg}1\,\unitenv{m}{-3}$ (yes, that is
$1\,\unitenv{g}1\,\unitenv{cm}{-3}$ because there are $100\,\unitenv{cm}1$
in $1\,\unitenv{m}1$ and you get to cube that to make
$1\,\unitenv{m}3$), so an absolutely full bucket of water has a mass
of nearly $30\,\unitenv{kg}1$ (that's really, really heavy).  An empty
bucket is much less massive, maybe even less than $1\,\unitenv{kg}1$,
but in detail it depends on whether it is a metal bucket or a plastic
one, and how thick are its walls.

What else can matter, physically, in the problem?  Clearly the
properties of the air must matter, because we have to assess whether
we are in the air-doesn't-matter regime!  Let's start with the
density.  Unfortunately, it is hard to estimate the density of the air
without some memory.\footnote{Actually, you can figure it out if you
  know that atmospheric pressure is about
  $15\,\unitenv{lb}1\,\unitenv{in}{-2}$ or
  $10^5\,\unitenv{N}1\,\unitenv{m}{-2}$ and that the scale-height of the
  atmosphere is about $10\,\unitenv{km}1$, but most students are more
  likely to remember secondary-school chemistry.}  You may recall from a
chemistry class that 1 mole of gas at standard temperature and
pressure (both close to the air temperatures and pressures you are
used to) fills a volume of about
$22\,\unitenv{\ell}1=0.022\,\unitenv{m}3$ and that the air is mainly
$\mathrm{N}_2$ gas with a molecular weight of about
$28\,\unitenv{g}1\,\unitenv{mol}{-1}$.  That makes air about
$1.3\,\unitenv{kg}1\,\unitenv{m}{-3}$ or about 1000 times less dense
than water.

In general, other properties of the air can matter to this, including
the viscosity.  We are going to cheat a bit by just assuming that it
is only the density of the air that matters here and not the viscous
properties; I will leave the demonstration of that as a problem for
the ambitious reader.  It also matters in detail whether there is a
strong wind.  For now we will assume not.

What else can matter, physically, in the problem?  We know that the
bucket will fall.  What makes it fall?  Gravity, of course!  So we
need to know about the gravitational force.  As I have noted, we all
know---not really from our general experiences (which are complex) but
rather from the textbook-physics classes we have taken (which are
over-simplified)---that ``all things fall at the same rate.''  What is
meant by this?  First of all, as I have said, there ought to be a
qualifier ``in the absence of air resistance'' and another ``in the
absence of buoyancy''; that is, the statement is only true when the
air doesn't matter, and gravity is the only important force.  But what
is meant be the word ``rate''?  Really the gravitational force sets
not the ``rates'' of falling objects (whatever they would be) but the
\emph{accelerations} of them.

There also ought to be another qualifier: ``when the two things are
near one another.''  Acceleration is a vector; it has a magnitude and
a direction.  Two objects close in space obtain the same acceleration
due to gravity, in both magnitude in direction.  But two bodies far
from one another will \emph{not} necessarily have the same
gravitational acceleration.  For example, bricks dropped in New York
and Paris fall with very similar acceleration \emph{magnitudes} but in
rather different acceleration \emph{directions}, since New York and
Paris are in different directions relative to the center of the Earth.
For another example, bricks dropped from sea level and from a
(theoretically possible) platform thousands of km above sea level---or
from a location inside a (theoretically possible) hole dug thousands
of km into the ground---will fall with different acceleration
magnitudes.  These bricks were dropped from locations that are
\emph{not} close in space.

The statement of Galilean gravity---that all things fall at the same
rate---ought to be---more precisely---``under conditions in which
gravity is the dominant force, all bodies that are close in space move
with the same acceleration.''

Whenever a word like ``close'' or ``far'' or ``big'' or ``small''
appears in a physics problem (or physics solution), it is necessary to
ask ``with respect to what?''.  When we say ``close in space'' in our
statement of the Galilean observation (which we will later come to
call the ``equivalence principle'' as Einstein and others promoted it
to a fundamental symmetry of the Universe), we mean ``close in space''
relative to what?  There is only one scale in the problem.  We mean
that the bodies are much closer to one another than they are to the
source of the gravitational acceleration which, in this case, is the
Earth.  If you think a bit longer about the New York and Paris
example, you will see that the scale is the distance to the
\emph{center} of the Earth and the condition is ``much closer together
than their distance to the center of the gravitating
body.''\footnote{It was not trivial for Newton to demonstrate,
  mathematically, that it is only this distance that matters, when the
  bodies are spherical, and not some more complicated distance computed
  from the shape of the body.  But even if he had found a more
  complicated relationship it would nonetheless be the case that the
  distance to the center of the Earth sets the approximate \emph{scale}
  of the problem, and the condition would be true at the
  order-of-magnitude level.}

Now, when the bucket falls from its third-story platform, it traverses
a distance $h$.  That distance (a matter of meters) is much, much
smaller than the distance to the center of the Earth (thousands of
kilometers).  Therefore, throughout its trajectory, the acceleration
of the bucket due to gravity is constant and the same.  It has
magnitude $\abs{\tv{g}}\equiv g\approx
10\,\unitenv{m}1\,\unitenv{s}{-2}$ and direction \emph{straight down.}

Do we have all of the possibly relevant physical quantities for this
problem?  I am not sure we do, but let's proceed.  One unfortunate
thing about the study of physics is that you \emph{never really know}
if you have all of the relevant physical effects under control, and
the history of the field is riddled with interesting examples of wrong
conclusions when the most relevant effect was ignored or unknown
(think ``perihelion precession of Mercury,'' which led to the
discovery of general relativity, or ``the energy source for the Sun,''
which was at one time thought to be gravitational, and caused Darwin
to doubt his theory of evolution\footnote{Although I am not a
  historian, it is clear from text in \textit{The Origin of the Species}
  that the contemporaneous physical understanding of the Sun made Darwin
  doubt his theory, because he could see that evolution must take place
  over very long times, much longer than the few million years the Sun
  would last if it were powered only by its own gravitational energy.}).
I will give you a rule of thumb for everyday mechanics however: If you
can account for gravity, and if you can account for everything that
mechanically \emph{touches} the body of interest, then you have
probably accounted for all of the forces.  We did gravity, and the
only thing that mechanically touches the bucket as it falls is the
air, the density of which we have estimated, so we satisfy this rule
of thumb in this case.  Of course all bets are off if you add electric
and magnetic fields, but these rarely matter in macroscopic mechanics
problems like this one.

Now, how do we solve the problem?

\section{Can we ignore air resistance?}

The first question is: Does the air matter, or can we ignore it?  It
turns out---as you may know---it is straightforward to solve this
problem if the air \emph{doesn't} matter.  Purely gravitational
trajectories are the stuff of textbooks.  We will perform this
calculation shortly.  But what to do if the air \emph{does} matter?
Since the bucket is more dense than the air (even if empty), you might
think that it is plausible to ignore air resistance and buoyancy.  But
of course a sheet of newspaper is much more dense than the air, and
yet it's trajectory can \emph{not} reasonably be approximated by
ignoring air resistance.

Trajectories of bodies in mechanics are set by considering
accelerations, and accelerations are set by forces (via the famous
equation $\tv{F}=m\,\tv{a}$.  (If that sentence is mysterious to you,
don't worry, we will come back to forces and accelerations again and
again in future lectures.)  So it makes sense for us to compare any air
resistance or buoyancy \emph{forces} to the \emph{force} of gravity.
By the famous law ``$\tv{F}=m\,\tv{a}$,'' the force on an object
accelerating with the acceleration $\tv{g}$ due to gravity---what you
might call the ``gravitational force'' $\tv{F}_{\mathrm{g}}$---is just
$\tv{F}_{\mathrm{g}}=m\,\tv{g}$.  We want to compare this force to the
forces of air resistance and buoyancy.  What are the magnitudes of
these forces?

Buoyancy is the simplest.  The buoyant force on the bucket has the
magnitude of the gravitational force on the air displaced by the
object, but the opposite direction.  One argument for this is that
parcels of still air neither rise nor fall because they have exactly
balanced gravitational and buoyant forces.  The buoyant force on an
object is a small correction to the gravitational force whenever the
object is much more dense than the air.\footnote{It sometimes comes as
a surprise that \emph{all} objects, including buckets and people and
automobiles, are subject to a buoyant force at all times.  Because
these bodies are not in a vacuum but rather surrounded by air, there
are energies and forces involved in the displacement of that air.
Your kitchen scale measures not the mass of your tomato, but the
\emph{difference} between the mass of the tomato and the mass of the
air it displaces.  Actually, it might be more complex than that,
depending on how the scale has been \emph{calibrated,} but that is a
subject for a later lecture.}  No problem there, for either the bucket
or the water it may contain.

Air resistance is more difficult, though not difficult.  We all have
the (correct) intuition that the air resistance force increases with
the object's speed and size.  We can attempt to estimate the magnitude
of the air resistance force by \emph{dimensional analysis}; that is,
we can find quantities that have the same \emph{dimensions} or
\emph{units} of force, depend in an increasing way on the object's
speed and size, and which relate somehow to the properties of the air
(of which we have only estimated one, its density).  We can then hope
that the quantities so found are close to what we are trying to
determine.

There is a miracle of physics that when you find a \emph{dimensionally
correct} answer, and when you have properly identified the most
important physical processes, you are almost always very close to the
\emph{exact} answer, as I will not argue, but as I will show by
example in what follows.  The ``dimensions'' of a physical quantity is
hard to define, but it is the part of it described by the units; for
example in this problem the height $h$ is measured in m or ft and
therefore has dimensions of \emph{length.}  The bucket has a property
measured in kg; it has dimensions of \emph{mass.}  The acceleration
due to gravity is measured in $\mathrm{m\,s^{-1}}$; it has dimensions
of \emph{length over time squared}.  In mechanics, all quantities have
dimensions which can be reduced to (possibly complex) combinations of
mass, Every correct expression in physics has correct dimensions; that
is, the dimensions on the left side of the equality must be the same
as those on the right side; this is an \emph{incredibly restrictive}
property of all physical expressions!

The technique of dimensional analysis---finding the only dimensionally
correct expression and then assuming or hoping that the exact
expression is close---is an incredibly powerful technique in physics.
We will use it in almost every lecture in this series.  It allows us
to rapidly explore hypotheses without doing a great deal of math or
analysis, and it provides very robust results.  The results aren't
\emph{precise} but they are \emph{robust} because although in general
the exact answer is unknown, every correct physical expression is
\emph{required} to have the correct dimensions; this ``symmetry'' is
not just restrictive but fundamental.\footnote{I am not sure it is
correct to say that dimensional analysis relies on a symmetry, but the
requirement of correct dimensions is very like a symmetry, in the way
that symmetries are used in physics.  For example, it is similar to
the restrictive requirement that if the left-hand side of an equation
is a vector expression, then the right-hand side must also be a vector
expression.}

Forces are described by the law $\tv{F}=m\,\tv{a}$ and therefore must
have dimensions of mass times acceleration.  An acceleration is a
length over a time squared.  So the dimensions of force are
\begin{equation}
\dimenv{force}1=\dimenv{mass}1\,\dimenv{acceleration}1
               =\dimenv{mass}1\,\dimenv{length}1\,\dimenv{time}{-2}
\quad .
\end{equation}
We expect the air resistance force to relate somehow to the speed of
the object, which has dimensions
\begin{equation}
\dimenv{speed}1=\dimenv{length}1\,\dimenv{time}{-1} \quad ,
\end{equation}
and to the density of air, which has dimensions
\begin{equation}
\dimenv{density}1=\dimenv{mass}1\,\dimenv{volume}{-1}
                 =\dimenv{mass}1\,\dimenv{length}{-3} \quad .
\end{equation}
To get \dimenv{mass}1 on top, and \dimenv{time}2 on the bottom, the
expression must involve density times speed \emph{squared.}  This
leaves the dimensions wrong by a factor of \dimenv{length}2, or area.
So there is a \emph{dimensionally correct} expression for the air
resistance force of the form
\begin{equation}
\dimenv{force}1=\dimenv{density}1\,\dimenv{area}1\,\dimenv{speed}{2}
\quad .
\end{equation}
It turns out that if we make the ``area'' the cross-sectional area of
the object---that is, a measure of the object's \emph{size}---and the
density the density of air, this is a reasonable expression for the
air resistance force!  It is wrong in detail, because there is a
\emph{dimensionless prefactor} of order unity that depends on the
specific shape and state of rotation of the object.  But right now we
are just trying to \emph{estimate} if there is \emph{any chance} that
air resistance matters in this problem.

Before we continue, allow me to remark\footnote{This remark is more
for the teachers of this material than for the students} that any book
or class that claims to ``solve'' a problem like this---a problem
relating to an object falling near the surface of the earth---in which
a quantitative analysis of the effect of the air has \emph{not} been
considered, has failed, utterly, to solve the problem.  This is
\emph{generic} in the textbook approach to physics.  The use of the
no-air equations is an exercise in mathematics, not physics; it is the
identification of the relevant physical effects that is the most
important work of physicists.  It certainly isn't physics if there is
no consideration or analysis or justification of the approximations.

Unfortunately, we don't yet know the \emph{speed} of the falling
bucket.  We have a chicken-and-egg problem here.  We can't calculate
the speed of the bucket properly if we can't ignore air resistance,
but we can't figure out whether we can ignore air resistance if we
can't calculate the speed of the bucket!  What to do?  We break the
impasse by ignoring air resistance and then \emph{evaluating,} after
the calculation, whether we were justified in doing so.\footnote{Since
air resistance will only \emph{slow} the fall, and the air resistance
force depends in an increasing way on speed, this one-sided approach
(assume air doesn't matter and then evaluate after) is safe.  There
are pathological situations where this will not work.}

If we ignore air resistance, what is the speed of the bucket?  Well,
it increases as it falls, because gravity sets the
\emph{acceleration}, not the speed.  It falls a distance $h$, a
measure of length, and gravity sets the acceleration $g$, a length per
time squared.  What is the speed of the bucket at its fastest point?
The only quantity with dimensions of speed in this problem is obtained
by taking the square root of the product, or
\begin{equation}
\sqrt{\dimenv{length}1\,\dimenv{acceleration}1}
 = \sqrt{\dimenv{length}2\,\dimenv{time}{-2}}
 = \dimenv{speed}1
\quad .
\end{equation}
This, up to a factor of order unity (a factor of $\sqrt{2}$ but it
just doesn't matter at this level of precision), is the speed of the
bucket when it hits the ground, if we ignore air resistance.

Putting it all together, we can safely ignore air resistance when the
magnitude $F_\mathrm{a}$ of the force from the air is much smaller
than the magnitude $F_\mathrm{g}$ of the force of gravity, or, in
dimensionless language, when the ratio is much smaller than unity.
Symbolically, we are okay when
\begin{equation}
\frac{F_\mathrm{a}}{F_\mathrm{g}}\approx\frac{\rho\,A\,v^2}{m\,g}
                                 \approx\frac{\rho\,A\,h}{m} \ll 1
\quad ,
\end{equation}
where $\rho$ is the density of the air, $A$ is the cross-sectional
area of the bucket as it falls, $h$ is the height, $m$ is the total
mass of the bucket and its contents, and we have used our previous
result (from dimensional analysis) that $v^2\approx g\,h$.  The astute
reader will notice that the gravitational acceleration $g$ has
factored out, and the limit is that in which the bucket is much more
massive than the \emph{column of air} through which it
falls!\footnote{See also Hogg, ``Air resistance,'' (arXiv:
physics/0609156), and Mahajan \& Hogg, ``Introductory physics: The new
scholasticism,'' (arXiv: physics/0412107).}

Can we ignore the air?  I have that $\rho\approx
1\,\unitenv{kg}1\,\unitenv{m}{-3}$, $A\approx 0.1\,\unitenv{m}2$,
$h\approx 10\,\unitenv{m}1$, and $1<m<30\,\unitenv{kg}1$ (depending on
how full the bucket is).  I estimated the area and maximum mass for
the bucket by assuming that it is about a cubic foot, and I assumed
$3\,\unitenv{m}1$ per story in the building.

Uh oh!  \emph{We can't ignore air resistance if the bucket is empty!}
The empty bucket gets a ratio $(F_\mathrm{a}/F_\mathrm{g})$ of unity.
The full bucket is fine (or fine enough for our precision today).  I
think this jives with our intuition or memory of similar events.
Imagine an empty bucket falling from the third story.  You can imagine
it getting slowed a bit by the air, approaching a constant speed,
maybe having its trajectory wander a bit left and right as it falls.
But a full, heavy bucket slams straight downwards, impervious to the
air or breeze.

\section{The solution and some discussion}

How to proceed?  For now let's simply note that the bucket must be
filled with water if we are going to safely ignore air resistance (we
have learned something!)\ but then assume that the bucket \emph{is}
filled with water and continue.  In this case, we can finally answer
the question posed at the beginning of this lecture: How long does the
bucket take to fall?

We seek a time; continuing with our dimensional analysis, in the
context of gravity, we have an acceleration $g$, a mass $m$ and a
height (length) $h$.  The only one of these that includes time at all
is $g$, and the only combination with dimensions of purely of time is
\begin{equation}
\sqrt{\dimenv{length}1\,\dimenv{acceleration}{-1}} = \dimenv{time}1
\quad .
\end{equation}
With $h\approx 10\,\unitenv{m}1$ and $g\approx
10\,\unitenv{m}1\,\unitenv{s}{-2}$, this \emph{characteristic time} is
$1\,\unitenv{s}1$.  And indeed, this probably agrees with your initial
guess based on your memory of similar incidents you have experienced.

This answer, $\sqrt{h/g}$, obtained by dimensional analysis, is in
fact off of an exact calculation by a factor of $\sqrt{2}$.  Do we
care?  Sure we do!  We \emph{do} have to get right answers sometimes.
This characteristic time answer is not quite correct, but it contains
almost all of the \emph{physics} in the problem.  Everything after
this is just math.  We just have to write down the correct equations
and solve them.  But notice that \emph{before we start} writing and
manipulating equations, we already know that our answer has to be some
unknown dimensionless constant times $\sqrt{h/g}$.  That's pretty
good.

The dimensional part of our approximate answer $\sqrt{h/g}$ is
correct, but more importantly, it is close---in an order-of-magnitude
sense---to the \emph{right} answer.  After all, we could have
manipulated numbers or symbols and obtained a characteristic time of
$100~\mathrm{ns}$, or we could have obtained $35~\mathrm{yr}$.  But in
fact when we wrote down the \emph{only} combination of things that we
could possibly have written down that could satisfy the dimensions
requirements, the quantity we wrote down is in fact close to correct!
That's significant.  We have got almost all of the way to the
complete, exact answer using \emph{nothing} but common sense and some
simple facts about gravity near the surface of the Earth.

But we learned more than this, we also learned something about the
\emph{scaling} of the answer with changes to the problem.  For
example, our answer (as we might have predicted) shows that the time
gets longer as the height increases (and \emph{how} it gets longer) or
as the acceleration due to gravity decreases (if, say, I saw this
happen on the Moon).  We also learned, without putting it in
explicitly, that the \emph{mass} $m$ of the bucket does \emph{not}
matter.  It matters to the estimation of the importance of air
resistance, but it does not matter to the free-fall time in the
absence of air resistance.  This, one of the main and most
counter-intuitive of Galileo's discoveries, falls naturally out of the
dimensional argument (which starts, of course, with Galileo's
observation that gravity sets the accelerations of falling objects to
$g$).

More precisely, what we have obtained from our dimensional argument is
the following: When the bucket is full and air can be ignored, the
free-fall time $t$ can be written in the form
\begin{equation}
t = Q\,\sqrt{\frac{h}{g}} \quad ,
\end{equation}
where $Q$ is a dimensionless prefactor.  All of the \emph{math} you
did when you solved this problem in textbook-physics classes was all
about just determining this one---relatively
uninteresting---dimensionless factor $Q$, which turns out to be
$\sqrt{2}$.

I have attempted to separate the physical reasoning part of the
problem from the mathematical details, and I have written here only
about the physical reasoning.  And although we did not get that
prefactor $Q$---we need to delve into the mathematical details to get
that---we did get a lot, like the scaling, and the order of magnitude.
We learned that I \emph{do} have time jump out of the way of the
falling bucket, but we also learned that I don't have time to make and
install a sign to warn others!  We also got lots of time to think
about physics and not have mathematical techniques interfere with
that.

We have some loose ends: What would we have done if the bucket were
empty?  Unfortunately, with a force---the air resistance ``ram
pressure'' force $\rho\,A\,v^2$---that goes like speed squared, there
is no analytic trajectory.  You have to \emph{integrate} the
trajectory with a computer.  Interestingly, this is very easy to do
with an accounting spreadsheet or any simple programming language or
data analysis package or anything equivalent, as we will see in later
lectures.  So we would not be \emph{doomed} if we were forced to
consider the empty bucket.  But our textbook physics would have failed
us, because in textbook physics you only consider problems that have
analytic solutions, which, as I have said, is a vanishingly small
minority of problems you might care about.

Another loose end we left was on the question of the accuracy with
which we estimate the height of the third story.  Our guess of
$10\,\unitenv{m}1$ is not good to better than, say,
$30\,\unitenv{percent}1$.  However, since the height appears in our
answer inside a square-root, a $30\,\unitenv{percent}1$ change to the
height leads to only about a $15\,\unitenv{percent}1$ change in the
free-fall time.\footnote{We will return to this kind of scaling
argument in future lectures.}  That uncertainty is smaller than our
missing factor of $\sqrt{2}$ and our uncertainties about the effects
of air resistance, and therefore not important at the level of
accuracy we have adopted.

That brings up a final note.  In this entire lecture, I have been
careless with the numbers and performed all estimation and mathematics
with little rigour.  I have found no precise or accurate answer to
anything.  This is not strange or unusual; this is absolutely
\emph{generic} for the study of physics.  One of the pervasive myths
about physicists is that we spend our time making very precise
predictions and measurements; after all, we know that the magnetic
moment of the electron is $1.001159652187\,\mu_B$.  But of course that
can happen only in a very specialized part of physics, it can be done
only in those rare circumstances in which we have an \emph{exact}
physical description with which we can exactly understand a physical
experiment and we have a precise experiment.  Only then do we perform
precise calculations and make precise measurements.  In most of the
physics we are going to be talking about---things like falling
buckets---there are a very large number of different physical effects
that are important if we want high precision.  Furthermore, many of
the effects are so complex that they are not computable at all!  So it
would be pointless in this problem to attempt to get a highly precise
answer for the time.  Generically, we will have to make approximations
and get an approximate answer.  This is another very important
distinction between the physics we do here and the textbook physics
you might find elsewhere.  \emph{We will not compute precise answers
when problems do not warrant them.}

\section{Problems}

\problem{From what story will the bucket take twice as long to fall
as it does from the third story?  Is air resistance important for that
fall, assuming the bucket is full?}

\problem{At what height $h$ does air resistance become important
for the free fall of a full bucket of water?}

\problem{A ball of mass $m$ is thrown upwards with speed $v$.  Use
dimensional analysis to estimate the height $h$ to which it will
ascend.  Ignore air resistance.  How wrong is your answer obtained by
dimensional analysis alone?}

\problem{A can throw with twice the velocity of B.  How much farther
can A throw than B, if you ignore air resistance?  Qualitatively, how
will your answer change if air resistance becomes important?}

\problem{Rank these objects in terms of the effect of air resistance
on their fall; that is, by how far they can fall before air resistance
becomes important: a penny, a brick, and a tennis ball.}

\problem{A great baseball outfielder can throw a baseball
$120\,\unitenv{m}1$.  Is that throw strongly affected by air
resistance?  Make the calculation by assuming that it isn't strongly
affected, compute the (approximate) speed of the throw, and then
assess.}

\problem{A professional golfer can hit a golf ball
$300\,\unitenv{m}1$.  Is that shot strongly affected by air
resistance?  Make the calculation by assuming that it isn't strongly
affected, compute the (approximate) speed of the shot, and then
assess.}

\problem{The air resistance force we computed in this lecture is a
kind of ``ram pressure.''  There is also a ``viscous drag'' term,
involving the viscosity of air.  Use a textbook value for the
viscosity of air and dimensional analysis to evaluate whether I was
justified in ignoring the viscous term relative to ram pressure for
the falling bucket.  What do you need to assume?}

\end{document}